\documentclass[twocolumn,aps,prl,nofootinbib,groupedaddress]{revtex4}
\usepackage{epsfig,amssymb,amsmath}
\usepackage[hypertex,linkcolor=red]{hyperref}

\usepackage{color}

\def\labell#1{\label{#1}}
\def\section#1{{\par\em #1:--- }}
\def\togli#1{}

\def\>{\rangle}
\def\<{\langle}
\begin{document}
%\fbox{{\scriptsize Eprint: quant-ph/}}
%\fbox{{\scriptsize Preliminary draft. \today}}
%\fbox{{\scriptsize Submitted paper.}}

%\fbox{{\scriptsize Final draft \today.}}
%Title of paper
\title{State estimation: direct state measurement vs.
  tomography} \author{Lorenzo Maccone, Cosimo
  C.~Rusconi} \affiliation{ \vbox{Dip.~Fisica ``A.~Volta'', INFN
    Sez.~Pavia, Univ.~of Pavia, via Bassi 6, I-27100 Pavia, Italy} }
\begin{abstract}
  We compare direct state measurement (DST or weak state tomography)
  to conventional state reconstruction (tomography) through accurate
  Monte-Carlo simulations. We show that DST is surprisingly robust to
  its inherent bias. We propose a method to estimate such bias (which
  introduces an unavoidable error in the reconstruction) from the
  experimental data. As expected we find that DST is much less precise
  than tomography. We consider both finite and infinite-dimensional
  states of the DST pointer, showing that they provide comparable
  reconstructions.
\end{abstract}
%\pacs{03.67.Lx, 03.67.Dd, 03.67.-a, 03.67.Ac}
%Quantum computation, 03.67.Lx
%Quantum cryptography, 03.67.Dd
%Quantum information, 03.67.-a
%quantum algorithms and protocols, 03.67.Ac
\maketitle

The no-cloning theorem prevents recovering the quantum state from a
single system \cite{yuen}. State reconstruction procedures must then
employ multiple copies of the system and are affected by statistical
errors: different procedures will have different efficiencies in
converging to the state. Here we compare two such procedures, the
recently introduced direct state measurement (DST) or
weak-value-tomography
\cite{lundeen,wiseman,altrisim,agarwal,holger,wu,boyd,finlandesi,yutaka},
and the well established quantum tomography
\cite{tomoreview,leonh,tomospin,quorum}. The DST is much less
demanding experimentally: it only entails (i) a weak coupling of the
system with an external pointer, (ii) a filtering (post-selection) of
the final state of the system and a simple projective measurement of
two complementary observables of the pointer.  Tomography, in
contrast, requires measuring a complete set of observables of the
system. However, the DST is less efficient than tomography and is a
biased procedure that introduces an unavoidable error in the
reconstructed state. Here we show that, surprisingly, DST is quite
robust to the bias, although it is known that it cannot reconstruct
arbitrary states \cite{finlandesi} and although (as expected) it is
much less efficient than tomography that can achieve similar precision
with several orders of magnitude less data. Moreover, we give a
prescription of how one can estimate the error introduced by the bias,
assuming that the state to reconstruct is pure: this allows the
experimenter to optimally tune the weak coupling to the pointer so
that such error is negligible with respect to (or comparable to) the
statistical error.  Finally, we show that the dimensionality of the
pointer system in DST is irrelevant: a two-level pointer (qubit) is as
efficient as an infinite-dimensional continuous-variable pointer
system.

We start by describing the theories behind DST and tomography. We then
show how these two methods compare through Monte-Carlo simulations of
reconstructions of physical systems of different dimensionality.

\section{Direct State Measurement} Two DST procedures exist: the first
assumes that the state to be reconstructed is pure, the second applies
to general (possibly mixed) states. In the first case, we choose a
basis $|n\>$ on which to perform the reconstruction and a state
$|c_0\>$ that is complementary to all: $|\<c_0|n\>|^2=1/d$ ($d$ being
the dimension of the system Hilbert space). In the second case, we
need a full complementary basis $|c_j\>$ such that
$\<c_j|n\>=\omega^{jn}/\sqrt{d}$, where $\omega\equiv e^{2\pi i/d}$.

The DST with qubit pointer \cite{lundeen,lundeenprl,wiseman} involves
weakly coupling the system to a pointer qubit through a unitary
coupling $U_{\varphi,n}=e^{-i\varphi|n\>\<n|\otimes \sigma_z}$
($\sigma_z$ the Pauli matrix) and then performing separate projection
measurements on the system and pointer\footnote{In place of
  $U_{\varphi,n}$ it is also possible to choose a coupling
  $e^{-i\varphi|n\>\<n|\otimes (\sigma_z-\theta|0\>\<0|)}$: to first
  order in $\varphi$, this coupling gives the same results as
  $U_{\varphi,n}$. A choice $\theta\neq 0$ is equivalent to
  introducing a phase factor $e^{-i\varphi\theta/2}$ to the $n$th
  component of the system state since
  $\sigma_z-\theta|0\>\<0|=(1-\tfrac\theta2)\sigma_z-\tfrac\theta2\openone$,
  but this phase factor is not advantageous, and we use $\theta=0$, as
  is customary in DST.}. The coupling is such that the pointer,
initially in an equal superposition $|+\>\equiv\tfrac
1{\sqrt{2}}(|0\>+|1\>)$, is rotated by an angle $\varphi$ if and only
if the system is in the state $|n\>$:
\begin{eqnarray}
  U_{\varphi,n}|n\>|+\>=|n\>(e^{-i\varphi}|0\>+e^{i\varphi}|1\>)/
\sqrt{2}.
\labell{qq}\;
\end{eqnarray}
After the
interaction, we measure both: on the pointer we measure the
expectation values of the Pauli matrices $\sigma_y$ and $\sigma_z$ (a
projective measurement on two sets of complementary bases); on the
system we either perform a simple post-selection on $|c_0\>$ (if we
know that the initial system state is pure) or a measurement on the
$|c_j\>$ basis in the general case.  Consider the latter: after the
projective measurement on the system with result $j$, the pointer
qubit is in the state
\begin{eqnarray}
&&\kappa^{(n)}_j\propto
\sum_{ml}\rho_{ml}\omega^{j(l-m)}\times\labell{asl}\\&&\nonumber
(e^{-i\varphi\delta_{mn}}|0\>+e^{i\varphi\delta_{mn}}|1\>)
(e^{i\varphi\delta_{ln}}\<0|+e^{-i\varphi\delta_{ln}}\<1|)
,\nonumber%\\&&\labell{asl}
\end{eqnarray}
where $\rho_{ml}=\<m|\rho|l\>$ is the initial state of the system in
the $|n\>$ basis and the Kronecker $\delta$ retains the memory of
which state $|n\>$ was coupled to the qubit. To first order in
$\varphi$, the expectation values of $\sigma_y$ and $\sigma_z$ on the
state $\kappa_j^{(n)}$ are proportional to the real and imaginary
value of $\sum_l\rho_{nl}\omega^{j(l-n)}$, namely
$\<\sigma_y\>+i\<\sigma_z\>\propto\sum_l\rho_{nl}\omega^{j(l-n)}$. We
can recover $\rho$ by inverting this relation:
\begin{eqnarray}
%\<\sigma_y\>+i\<\sigma_z\>\propto
%\sum_l\rho_{nl}\omega^{j(l-n)}\Rightarrow
\rho_{nq}\propto\sum_j\omega^{j(n-q)}\:
\mbox{Tr}[\kappa^{(n)}_j(\sigma_y+i\sigma_z)]
%[\<\sigma_y\>+i\<\sigma_z\>]
\labell{rh}\;,
\end{eqnarray}
where we recall that $\omega^{j(l-n)}=\<l|c_j\>\<c_j|n\>$.  Equation
\eqref{rh}, valid only to first order in $\varphi$, yields the state
of the system $\rho_{nq}$ from the experimental $\sigma_y$, $\sigma_z$
expectation values of the pointer by renormalizing it so that
$\sum_n\rho_{nn}=1$.

If the initial state of the system is pure $\rho=|\psi\>\<\psi|$, a
post-selection on $|c_0\>$ is sufficient: in fact, to first order in
$\varphi$ we have
\begin{eqnarray}
\mbox{Tr}[\kappa^{(n)}_0(\sigma_y+i\sigma_z)]%=\<\sigma_y\>+i\<\sigma_z\>
\propto\sum_l\<n|\psi\>\<\psi|l\>\propto\<n|\psi\>
\labell{pure}\;,
\end{eqnarray}
whence the state $\psi_n=\<n|\psi\>$ can be recovered upon
renormalizing Eq.~\eqref{pure} so that $\sum_n|\psi_n|^2=1$.

Up to now we have considered a pointer qubit, but analogous relations
hold if one uses a continuous-variable pointer \cite{boyd,wu,agarwal,cinesi}
initially in a Gaussian state $|\phi_{ptr}\>\propto\int dx\:
e^{-x^2}|x\>$. In this case, the interaction with the system shifts
the pointer position by $\varphi$ iff the system is in the state
$|n\>$, through a unitary $e^{i\varphi |n\>\<n|\otimes
  P}$ where $P$ is the pointer's momentum operator. After a
measurement of $|c_j\>$ on the system, the pointer is left in the
state
\begin{eqnarray}
\lambda_j^{(n)}\!\propto\!\sum_{ml}\rho_{ml}\omega^{j(l-m)}
\!\int\! dxdy\:e^{-(x+\varphi\delta_{mn})^2-(y+\varphi\delta_{ln})^2}|x\>\<y|
.\;\nonumber
\end{eqnarray}
To first order in $\varphi$, the expectation value of the position $X$
and momentum $P$ of the pointer are proportional to the real and
imaginary value of $\sum_l\rho_{nl}\omega^{j(l-n)}$, so we can recover
the state by inverting this relation:
\begin{eqnarray}
\rho_{nq}\propto\sum_j\omega^{j(n-q)}\mbox{Tr}[\lambda_j^{(n)}(X+iP)]
\labell{rhog}\;,
\end{eqnarray}
which (to first order in $\varphi$) gives the state of the system in
terms of the expectation values of $X$, $P$ of the pointer.

All the above procedures, derived to first order in $\varphi$, are
correct only in the limit $\varphi\to0$.  In (real and simulated)
experiments, one has to use a finite nonzero $\varphi$: this is a bias
that introduces an unavoidable error in the DST reconstruction (we
will show how to estimate it from the experimental data). Note also
that in the expansion that gives rise to Eq.~\eqref{pure}, the first
order term is proportional to $\<c_0|\psi\>^{-1}$: if the overlap of
the unknown state $|\psi\>$ to be reconstructed and the post-selected
one $|c_0\>$ is small, the first order approximation may fail
\cite{suda,cinesi}. In particular, it will be impossible to reconstruct a
state that is orthogonal to the post-selected one \cite{finlandesi}.
In practice it might be necessary to repeat the reconstruction using a
couple of different orthogonal post-selected states (e.g.~$|c_0\>$ and
$|c_1\>$), choosing the one that yields the least statistical
fluctuations in the reconstructed state. The formulas for general
states, Eqs.~\eqref{rh}, \eqref{rhog}, are immune to this
problem.

\section{Tomography} Tomography \cite{tomoreview} is a
well-established unbiased state reconstruction technique, affected by
statistical errors only. It can be understood very simply by
considering the system operators as a Hilbert space and choosing a
basis $B(\lambda)$ for it. Then any operator $A$ can be expanded on such a
basis as \cite{quorum}
\begin{eqnarray}
A=\int d\lambda\: B(\lambda)\:\mbox{Tr}[B^\dag (\lambda) A]
\labell{hs}\;,
\end{eqnarray}
where Tr$[B^\dag (\lambda) A]$ is the Hilbert-Schmidt scalar product between
operators $B(\lambda)$ and $A$. Now, set $A=\rho$, the state of the system,
and choose a basis $B(\lambda)=f(\lambda,O_\lambda)$ which is a function of an
observable $O_\lambda$ for all $\lambda$. We can recover the state $\rho$ using
Eq.~\eqref{hs}, since the trace there is just a function of the
outcome probabilities of measurements of these observables:
\begin{eqnarray}
\rho=\int d\lambda\:B(\lambda)\sum_jf\left(\lambda,o^{\lambda}_j\right)\<o_j^{\lambda}|\rho|o_j^{\lambda}\>
\labell{tomo}\;,
\end{eqnarray}
where $o_j^{\lambda}$ and $|o_j^{\lambda}\>$ are the eigenvalues and eigenvectors
of $O_\lambda$. The simplest example is the tomography of a qubit
\cite{tomospin}, where one can choose the Pauli matrices
$\openone,\sigma_x,\sigma_y,\sigma_z$ as a basis set $B(\lambda)$, so that
Eq.~\eqref{tomo} becomes
\begin{eqnarray}
\rho=\tfrac 12+\sum_{\gamma=x,y,z}\sigma_\gamma\sum_{m=-1/2,1/2}
m\:p(m|\sigma_\gamma)
\labell{tomosp}\;,
\end{eqnarray}
where $p(m|\sigma_\gamma)$ is the probability of obtaining the
eigenvalue $m$ when measuring $\sigma_\gamma$. For qudits with
arbitrary dimension $d$ one can obtain similar relations from
irreducible unitary representations of the $SU(2)$ group
\cite{tomospin}.

%da plot-weaktomo2.f che usa dati da qdit-tomo.f (fatto girare due
%volte e rinominando weaktomo.dat in weaktomo2.dat) e qdit-weaktomo.f
%parametri per grafico sx: alpha=2,N=500,NBLK=100,DDELTA=2.7,phi=0.05
%e phi=0.1, d=2 (v. directory fig1a)
%parametri per grafico dx: alpha=2,N=500,NBLK=100,DDELTA=2.6,phi=0.05
%e phi=0.1, d=10 (v.directory fig1b)
\begin{figure}[hbt]
\begin{center}
\epsfxsize=1.\hsize\leavevmode\epsffile{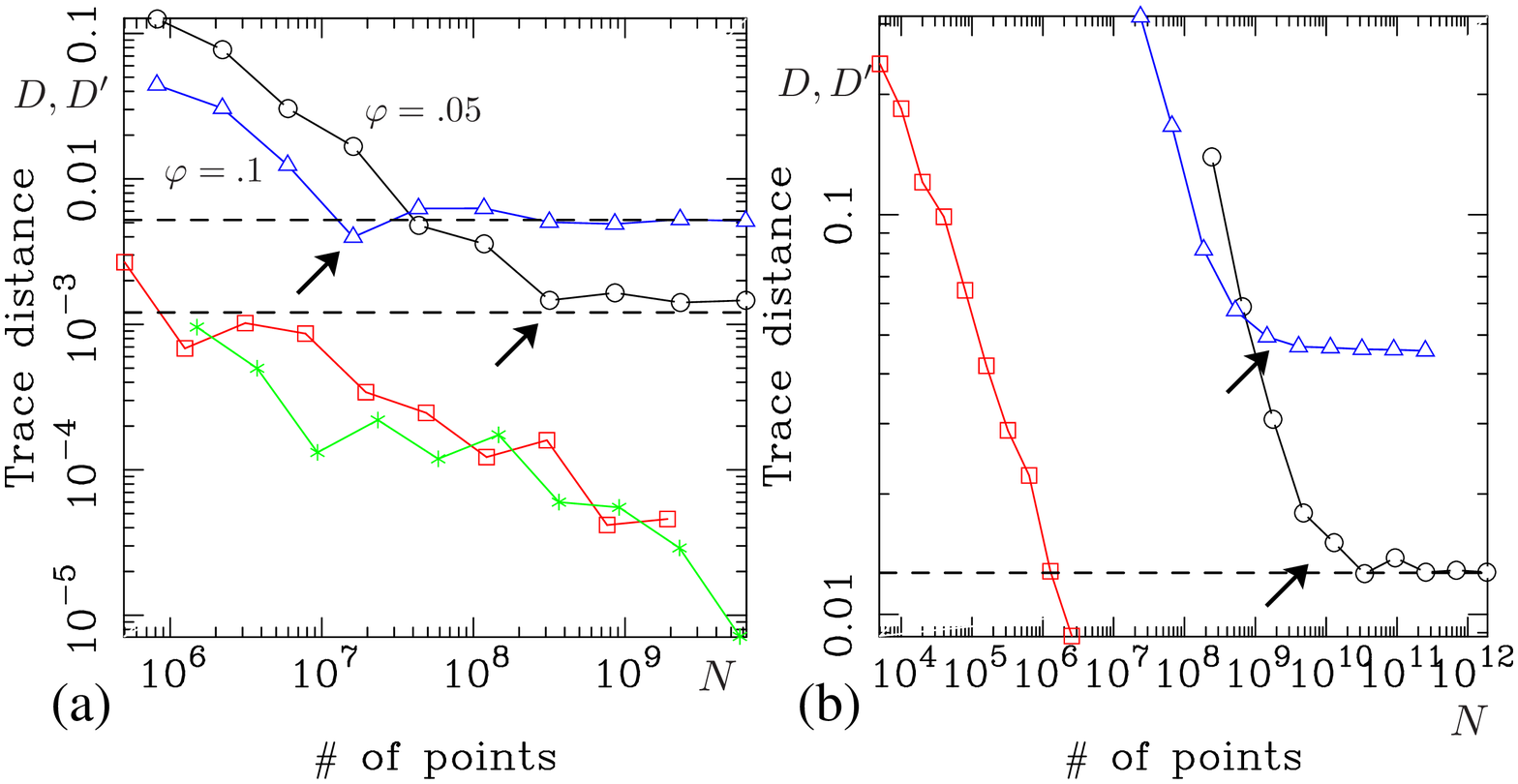}
\end{center}
\vspace{-.5cm}
\caption{Comparison between pure-state DST \eqref{pure} and tomography
  \eqref{tomo} for qubits (left) and for qdits (right).  The trace
  distance $D$ between the reconstructed $\rho^r$ and the true
  $\rho^t$ states is plotted as a function of the copies $N$ of the
  system state employed.  (a)~Circles: $D$ calculated from DST with
  $\varphi=0.05$; triangles: $D$ calculated from DST with
  $\varphi=0.1$; stars: $D$ using tomography with a basis of Pauli
  matrices through \eqref{tomosp}; squares: $D$ using tomography with
  a continuous basis taken from a unitary representation of $SU(2)$
  (as evident from the graph, these two bases are equivalent for
  tomography). As expected, tomography has smaller trace distance than
  DST. Moreover, DST saturates to the bias, indicated by the
  horizontal dashed lines at $D'$ of Eq.~\eqref{dprimo}, estimated
  from the experimental data: the arrows indicate the locations of the
  elbows where the trace distance starts to saturate as a function of
  $N$ (they correspond to the optimal number of measurements for the
  $\varphi$ employed). Here we simulate a qubit spin-coherent state
  \cite{tomospin} with parameter $\alpha=2$.  (b)~Same as previous
  ($\varphi=0.05,0.1$) but for a qdit with $d=10$. The post-selection
  of the pure-state DST drastically reduces the efficiency over
  tomography. Here we simulate a $d=10$ spin-coherent state with
  $\alpha=2$.}
\labell{f:fig1}\end{figure}
\section{Comparison} We present a comparison between DST and
tomography through simulated experiments. To gauge the quality of the
reconstruction we use the trace distance \cite{fuchs}
$D(\rho^{t},\rho^r)=$Tr$(|\rho^t-\rho^r|)/2$ between the true state
$\rho^t$ and the reconstructed one $\rho^r$: it gives the probability
of error in discriminating among the two, $p_{err}=1/2(1-D)$, and is a
number in the interval $[0,1]$. Even though this quantity is not
accessible in experiments (as typically one does not know the true
state $\rho^t$), for the pure-state case it can be estimated from the
data (see below).

%v directory fig2a,fig2b
%usa qdit-weaktomo-mixed.f
\begin{figure}[hbt]
\begin{center}
\epsfxsize=1.\hsize\leavevmode\epsffile{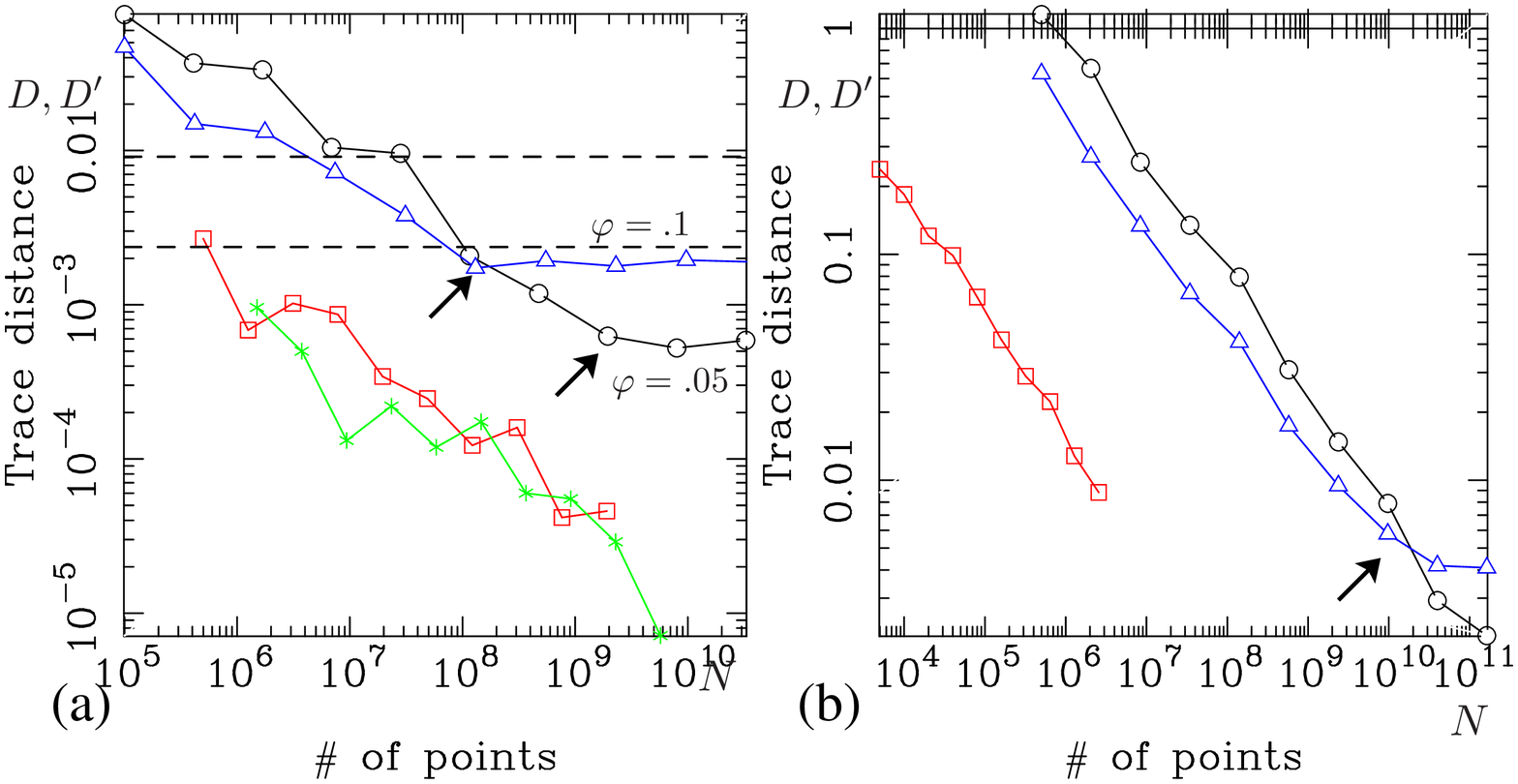}
\end{center}
\vspace{-.5cm}
\caption{Same as Fig.~\ref{f:fig1}, but using the general DST
  \eqref{rh}, with no post-selection: all data is employed in the
  reconstruction. While there is no substantial gain in efficiency
  over the pure DST \eqref{pure} for qubits (left), there is a
  substantial gain for qdits $d=10$ (right). Note that the trace
  distance can exceed 1 because the statistical fluctuations in the
  reconstructed state $\rho^r$ through Eq.~\eqref{rh} may yield an
  invalid quantum state (with negative eigenvalues).  The bias is
  estimated (horizontal dashed lines) using the formulas for pure
  state reconstruction \eqref{dprimo}, which overestimate the bias of
  general DST (it has only indicative value here).}
\labell{f:mixed}\end{figure}

Both DST and tomography are affected by statistical errors: in (real
and simulated) experiments one must estimate the expectation values in
Eqs.~\eqref{rh}, \eqref{pure}, \eqref{rhog}, and \eqref{tomo} from a
finite number $N$ of measurement outcomes. Of course, the great
advantage of DST is its incredible experimental simplicity but, as
expected, tomography is much more efficient: using the same number $N$
of measurements it can achieve a smaller trace distance than what can
be achieved through pure-state DST of Eq.~\eqref{pure}
(Fig.~\ref{f:fig1}) or general DST of Eq.~\eqref{rh}
(Fig.~\ref{f:mixed}).  Namely, achieving precisions comparable to
tomography with DST requires orders of magnitude more data. Moreover,
the precision of pure-state DST has a very strong dependence on the
overlap between the post-selected state $|c_0\>$ and the (unknown)
state to be reconstructed \cite{finlandesi}. In addition, DST is a
biased technique due to the finiteness of $\varphi$.  Even increasing
$N$, the trace distance $D$ will saturate to a nonzero value (see
Figs.~\ref{f:fig1}, \ref{f:mixed}): even for infinite statistics
$N\to\infty$ the wrong state is reconstructed $\rho^{t}\to\rho^a$,
where $\rho^a$ is the state obtained from \eqref{rh} using the ``true
state'' $\kappa_j^{(n)}$ instead of using only its first order
expansion in $\varphi$.  This bias implies that (in contrast to
tomography) in DST there is an optimal number $N$ of measurements that
should be performed for each $\varphi$, namely the $N$ at which the
trace distance $D$ starts to saturate (black arrows in
Figs.~\ref{f:fig1}, \ref{f:mixed}): for smaller $N$ the trace distance
has large statistical fluctuations (Fig.~\ref{f:errors}), while for
larger $N$ the trace distance does not decrease.  This optimal $N$
depends on $\varphi$ (larger $\varphi$ require smaller $N$ to
saturate) and on the unknown state, so it is unfortunately impossible
to predetermine it before the experiment.  Nonetheless, one can still
verify (at least qualitatively) when this number has been reached by
estimating $D$.

%Figura sx preparata da Cosimo. Figura dx preparata con
%plot-weaktomo2.f con i batch di dir fig1a.
\begin{figure}[hbt]
\begin{center}
\epsfxsize=1.\hsize\leavevmode\epsffile{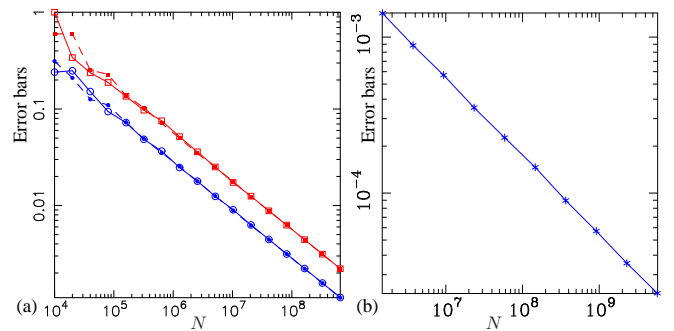}
\end{center}
\vspace{-.5cm}
\caption{Comparison of the error bars of DST (left) and tomography
  (right), with the same parameters of Fig.~\ref{f:fig1}a. The central
  limit theorem ensures a $\sim 1/\sqrt{N}$ scaling of the error bars.
  (a)~Statistical error bars of the DST reconstruction for qubits. The
  upper lines (squares) refer to $\varphi=0.05$, the lower lines
  (circles) to $\varphi=0.1$: a smaller $\varphi$ gives a better
  quality reconstruction, but higher statistical fluctuations. The
  dashed lines refer to a qubit pointer, the continuous ones to a
  Gaussian pointer (the two have substantially identical performance).
  (b)~Statistical error bars for the Pauli tomography of
  Eq.~\eqref{tomosp}.  The fluctuations are orders of magnitude
  smaller than DST, showing its larger efficiency.  }
\labell{f:errors}\end{figure}

We can estimate $D$ from the experimental data \cite{noi} without any
knowledge of the true state $\rho^t$ in the case of pure state DST
\eqref{pure} (see Fig.~\ref{f:vs}).  This can be done iteratively
using $D'=$Tr$(|\rho^e-\rho^r|)/2\simeq D$, where
$\rho^r=|\psi^r\>\<\psi^r|$ is the reconstructed state obtained from
the experimental expectation values of $\sigma_y,\sigma_z$ in
\eqref{pure}, while $\rho^e=|\psi^e\>\<\psi^e|$ is the `estimated true
state' obtained by analytically evaluating such expectation values on
the state $\sum_l{\psi^r}_l
(e^{-i\varphi\delta_{ln}}|0\>+e^{i\varphi\delta_{ln}}|1\>)/{\cal N}$
obtained from \eqref{pure} without first order approximation ($\cal N$
the norm and ${\psi^r}_l=\<l|\psi^r\>$ the components of $|\psi^r\>$).
In other words, we estimate $D(\rho^a,\rho^{t})$ iteratively from
$D'=D(\rho^e,\rho^r)$ where $\rho^r\to\rho^a=|\psi^a\>\<\psi^a|$ in
the limit of infinite statistics, and where the same mathematical
procedure gives $\rho^a$ from $\rho^t$ or $\rho^e$ from $\rho^a$.  The
estimate $D'$ converges very quickly to $D$ %(see
%Fig.~\ref{f:bias}) 
because, in the limit of infinite statistics $N\to\infty$,
$|\<\psi^t|\psi^{r}\>|^2=|\<\psi^r|\psi^{e}\>|^2+o(\varphi^5)$ where
the orders 1-3 in the $\varphi$-expansion are null.  For finite $N$,
$D'$ qualitatively approximates $D$ for the physically significant
small values of the parameter $\varphi$ (Fig.~\ref{f:vs}).  The
explicit form of $D'$ is
\begin{eqnarray}
D'=\Big[1-\Big|\tfrac1{{\cal
    N}^2}\sum_n{\psi^r}_n\sum_{ml}{\psi^r}_m({\psi^r}_l)^*
\gamma_{ml}^n\Big|^2\Big]^{1/2},
\labell{dprimo}
\end{eqnarray}
where $\gamma_{ml}^n\equiv
2\sin[\varphi(\delta_{mn}+\delta_{ln})]-2\sin[\varphi(\delta_{mn}-\delta_{ln})]$
and we used the formula $(1-|\<\chi|\theta\>|^2)^{1/2}$ for the trace
distance between pure states $|\chi\>,|\theta\>$. This expression only
uses the experimentally accessible data ${\psi^r}_l$. [An analogous
procedure applies to the Gaussian pointer, using Eq.~\eqref{rhog}.] A
formula similar to \eqref{dprimo} for the general DST of
Eq.~\eqref{rh} is currently lacking.

%Figure preparate da Cosimo
\begin{figure}[hbt]
\begin{center}
\epsfxsize=1.\hsize\leavevmode\epsffile{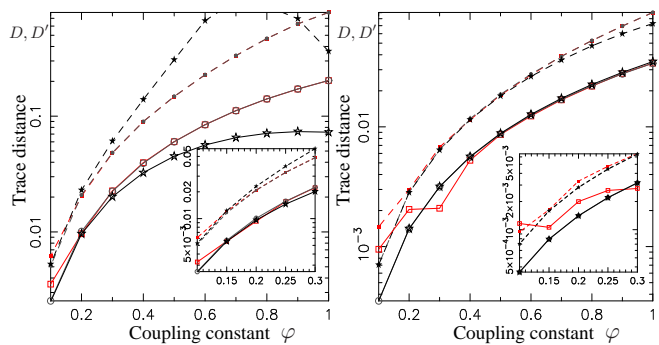}
\end{center}
\vspace{-.5cm}
\caption{Estimation of the trace distance from the experimental data
  for the qubit pointer (dashed lines, filled symbols) and the
  Gaussian pointer (continuous lines, empty symbols) as a function of
  the coupling $\varphi$. squares: trace distance
  $D(\rho^{t},\rho^r)$; stars: experimentally accessible trace
  distance $D'=D(\rho^r,\rho^e)$; circles: trace distance
  $D(\rho^{t},\rho^a)$.  All these trace distances approximate one
  another in the physically significant region of small coupling
  $\varphi$ (see magnification in the inset, where triangles and
  circles are indistinguishable), so we can approximate $D$ with the
  experimentally accessible $D'$.  Moreover, the substantial overlap
  of the curves with empty and filled stars shows that using a
  Gaussian or a qubit pointer is equivalent. Here we simulate the
  reconstruction of a qubit coherent state with $\alpha=\pm 2$
  ($-$~left with $N=49\times 10^6$ total repetitions, $+$~right with
  $N=159\times 10^6$).}
\labell{f:vs}\end{figure}

A comparison between the discrete pointer in DST of
Eqs.~\eqref{rh},\eqref{pure} and the continuous variable one of
\eqref{rhog} shows that the two reconstructions are substantially
equivalent for physically significant small values of $\varphi$ (see
Fig.~\ref{f:vs}). In closing, we note that by reconstructing the state
$|\psi^r\>$ with different values of $\varphi$, we can extrapolate it
to $\varphi\to 0$, in order to (partially) overcome the intrinsic bias
of DST \cite{noi}. 
\section{Conclusions}
We compared DST and conventional tomography through Monte-Carlo
simulated experiments. Our results: (1)~DST is surprisingly robust to
its inherent bias, due to the finiteness of the coupling $\varphi$ in
practice; (2)~a prescription of how one can estimate this (unknown)
bias from the experimental data only; (3)~the dimensionality of the
pointer system is irrelevant: a qubit or continuous-variable pointers
achieve the same performance; (4)~as expected, conventional tomography
achieves better performance for the same number of data, (but is
experimentally more complicated).

We thank Prof.~H.F. Hofmann for useful discussions.

\end{document}